\documentclass[conference]{IEEEtran}
\IEEEoverridecommandlockouts
\usepackage{cite}
\usepackage{amsmath,amssymb,amsfonts}
\usepackage{algorithmic}
\usepackage{graphicx}
\usepackage{textcomp}
\usepackage{xcolor}
\def\BibTeX{{\rm B\kern-.05em{\sc i\kern-.025em b}\kern-.08em
    T\kern-.1667em\lower.7ex\hbox{E}\kern-.125emX}}
\begin{document}

\title{Deep learning-based filtering of cross-spectral matrices using generative adversarial networks
\thanks{This research has been funded by German Federal Ministry for Economic Affairs and Climate Action
	(Bundesministerium für Wirtschaft und Klimaschutz BMWK) under project \emph{AntiLerM} registration number 49VF220063.}
}

\author{\IEEEauthorblockN{Christof Puhle}
\IEEEauthorblockA{\textit{Department of Signal Processing} \\
\textit{Society for the Advancement of Applied Computer Science (GFaI) e.V.}\\
Berlin, Germany\\
puhle@gfai.de}
}

\maketitle

\begin{abstract}
In this paper, we present a deep-learning method to filter out effects such as ambient noise, reflections, or source directivity from microphone array data represented as cross-spectral matrices. Specifically, we focus on a generative adversarial network (GAN) architecture designed to transform fixed-size cross-spectral matrices. Theses models were trained using sound pressure simulations of varying complexity developed for this purpose. Based on the results from applying these methods in a hyperparameter optimization of an auto-encoding task, we trained the optimized model to perform five distinct transformation tasks derived from different complexities inherent in our sound pressure simulations.
\end{abstract}


\section{Introduction}
As extensively investigated in \cite{Grumiaux2022}, state-of-the-art deep-learning methods for acoustical sound source localization (SSL) aim to directly reconstruct the direction of arrival of sources or, more generally, the parameters describing the acoustic scene in the presence of reverberation or diffuse noise. This article addresses the problem from a different perspective by employing generative adversarial networks (GANs) to either remove or, at least, reduce the effects of ambient noise, reflections, or source directivity in microphone array data (that is, cross-spectral matrices) before any potential SSL analysis begins. On the one hand, this approach improves the starting point for solving the SSL problem, on the other hand, it enables a more effective use of traditional mapping methods such as standard beamforming or CLEAN-SC.

This paper proceeds by first outlining our sound pressure simulation approach in \ref{simulations}, then presenting the machine learning model (see \ref{m-l-model}) that we designed to filter simulated cross-spectral matrices. Finally, we discuss results for five different transformation or filtering tasks in \ref{results}.

\section{Acoustic simulations}\label{simulations}
\subsection{Basics}
Let $\bar{p}:\mathbb{R}^3\rightarrow\overline{\mathbb{C}}$ be the complex amplitude of a time-harmonic sound pressure field of angular frequency $\omega>0$ and speed of propagation $c>0$. By definition, $\bar{p}$ satisfies the Helmholtz equation
\begin{equation}
	\frac{\partial^2\bar{p}}{\partial x^2}+\frac{\partial^2\bar{p}}{\partial y^2}+\frac{\partial^2\bar{p}}{\partial z^2}+k^2\bar{p}=0,
	\quad k=\frac{\omega}{c}.
\end{equation}
Our sign convention for a time-harmonic function is $t\mapsto\exp(i\omega t)$. For example,
\begin{equation}
	\bar{p}(x,y,z)=\frac{\exp\left(-ik\sqrt{x^2+y^2+z^2}\right)}{\sqrt{x^2+y^2+z^2}}
\end{equation}
represents an outgoing spherical wave with source in $(0,0,0)$.

Let $p:[0,\infty]\times[0,\pi]\times[0,2\pi]\rightarrow\overline{\mathbb{C}}$ be $\bar{p}$'s representation
in spherical coordinates $(r,\theta,\phi)$. Solutions to the corresponding Helmholtz equation can be found analytically
by assuming that ${p}$ is separable, i.e.\ there exist
functions $R:[0,\infty]\rightarrow\overline{\mathbb{C}}$, $\Theta:[0,\pi]\rightarrow\overline{\mathbb{C}}$,
$\Phi:[0,2\pi]\rightarrow\overline{\mathbb{C}}$
such that
\begin{equation}
	{p}(r,\theta,\phi)=R(r)\cdot\Theta(\theta)\cdot\Phi(\phi).
\end{equation}
In this case, the Helmholtz equation leads necessarily to
\begin{equation}
	R(r)=A\cdot h^{(1)}_l(kr) + B\cdot h^{(2)}_l(kr)
\end{equation}
for some constants $A,B\in\mathbb{C}$, $l\in\mathbb{N}_0=\{0,1,\ldots\}$, and $h^{(1)}_l$, $h^{(2)}_l$ denote the
spherical Hankel functions of the first and second kind of degree $l$, respectively. Moreover, we have
\begin{equation}
	\Theta(\theta)\cdot\Phi(\phi)=Y^m_l(\theta,\phi),
\end{equation}
where $m\in\{-l,-l+1,\ldots,l\}$, and $Y^m_l$ is the spherical harmonic of degree $l$ and order $m$.

\subsection{Smooth spherical pistons}\label{AA}

A vibrating spherical cap piston with aperture angle $\alpha\in(0,\pi]$ centered on the north pole of an otherwise rigid sphere with radius $r_0>0$ can be described by its surface velocity $v^{\alpha}:[0,\pi]\times[0,2\pi]\rightarrow\mathbb{R}$,
\begin{equation}
	v^{\alpha}(\theta,\phi)=V\cdot a^{\alpha}(\theta,\phi), \quad V>0,
\end{equation}
the corresponding aperture function $a^{\alpha}:[0,\pi]\times[0,2\pi]\rightarrow\mathbb{R}$ is given by
\begin{equation}
	a^{\alpha}(\theta,\phi)=1-H\left(\theta-\frac{\alpha}{2}\right), \quad H(x)=\begin{cases}
		0 & x<0\\
		1 & x\geq 0
	\end{cases}.
\end{equation}
The spherical wave spectrum of $v^{\alpha}$,
\begin{equation}\label{SpherWavSpecOfVelo}
	v^{\alpha}(\theta,\phi)=\sum_{l=0}^{\infty}\sum_{m=-l}^{l}v^{\alpha}_{lm}\cdot Y^m_l(\theta,\phi),
\end{equation}
can be computed via integration of the corresponding associated Legendre polynomials:
\begin{equation}
	v^{\alpha}_{lm}=V\delta_{m0}\sqrt{(2l+1)\pi}\int_{\cos\left(\frac{\alpha}{2}\right)}^1 P_l^0(x)dx.
\end{equation}
Rotating the spherical cap to be centered in the direction $(\tilde{\theta},\tilde{\phi})$ results in
\begin{equation}\label{SpherCoefRot}
	\tilde{v}^{\alpha}_{lm}=\sqrt{\frac{4\pi}{2l+1}}Y_l^m(\tilde{\theta},\tilde{\phi})^{\ast}\cdot v^{\alpha}_{lm},
\end{equation}
respectively. Finally, the radiated pressure in the region $r>r_0$ is completely determined by the surface velocity spectrum (see for example \cite{Williams1999}):
\begin{equation}\label{SpherCapPress}
	{p}_{v^{\alpha}}(r,\theta,\phi)=-i\rho_0c\sum_{l=0}^{\infty}\sum_{m=-l}^{l}{\frac{h^{(2)}_l(k r)}{h^{(2)\,\prime}_l(k r_0)}v^{\alpha}_{lm}\cdot Y^m_l(\theta,\phi)}.
\end{equation}

As most of the higher degrees in \eqref{SpherWavSpecOfVelo} are present to form the discontinuity at the boundary of the spherical cap, we opt for a smooth one-parameter family of spherical pistons as fundamental building block of our acoustical models. Its surface velocity $w^{\alpha}:[0,\pi]\times[0,2\pi]\rightarrow\mathbb{R}$ is defined via
\begin{equation}
	w^{\alpha}(\theta,\phi)=\begin{cases}
		V\cdot\exp\left(-\frac{\left(1-\cos\left(\theta\right)\right)^2}{\left(1-\cos\left(\frac{\alpha}{2}\right)\right)^2}\right) & \alpha\leq\pi,\\
		\frac{2\pi-\alpha}{\pi}\cdot w^{\pi}(\theta,\phi) + \frac{\alpha-\pi}{\pi}\cdot V & \pi<\alpha\leq 2\pi.
	\end{cases}
\end{equation}
Again, we will call $\alpha$ the aperture angle of this piston, but now, when varying $\theta$ from $0$ to $\alpha/2$, the particle velocity smoothly changes from $V$ to $V/e$ in the case $\alpha\leq\pi$. As before, the spherical wave spectrum of $w^{\alpha}$ can be determined by one-dimensional integration,
\begin{equation}
	w^{\alpha}_{lm}=\delta_{m0}\sqrt{(2l+1)\pi}\int_{-1}^{1} P_l^0(x)\hat{w}^{\alpha}(x)dx,
\end{equation}
where $\hat{w}^{\alpha}(\cos\left(\theta\right))\equiv w^{\alpha}(\theta,\phi)$. Moreover, transformation rule \eqref{SpherCoefRot} also holds for the coefficients $w^{\alpha}_{lm}$ when rotating the smooth spherical piston to be centered in the direction $(\tilde{\theta},\tilde{\phi})$, and the radiated sound pressure ${p}_{w^{\alpha}}$ corresponding to $w^{\alpha}$ can be computed in complete analogy to \eqref{SpherCapPress}. 

\subsection{Acoustic models}\label{acousticModels}

The simulations involved a set of $5000$ acoustic models fixed beforehand and each consisting of three (outgoing) smooth spherical pistons which were rotated and translated uniformly at random along the plane $z=0$ within a cube of edge length $2.56\,\textrm{m}$ that is centered at the origin. Moreover, each source is furnished with its own reflection plane together with a reflection coefficient between $-3\,\textrm{dB}$ and $-15 \,\textrm{dB}$. The aperture angles of the pistons vary from $3\pi/2$ to $2\pi$ (acoustic monopole) and source radii $r_0$ are chosen randomly and uniformly between $0.1\,\textrm{m}$ and $0.3\,\textrm{m}$. Within each model, the source $V$'s are chosen to be at most $15\,\textrm{dB}$ below the model sound velocity level, which ranges uniformly between $35\,\textrm{dB}$ and $85 \,\textrm{dB}$ across the model set. Consequently, the maximum dynamic range between sources within each model is $15\,\textrm{dB}$. Model temperatures are taken from a normal distribution with a mean of $20\,\textrm{C}^\circ$ and a standard deviation of $2.5\,\textrm{C}^\circ$. 

We approximated the sound pressures of these models up to Helmholtz degree $15$ at the positions of a virtual microphone array of spherical shape ($48$ microphones, diameter $0.35\,\textrm{m}$, centered at $(0,0,d)$ with $d=2.56\,\textrm{m}$) for $16$ distinct frequencies:
\begin{equation}
	10\cdot\Delta_f,\ldots,25\cdot \Delta_f,\quad \Delta_f = \frac{192000}{1024}\,\textrm{Hz}=187.5\,\textrm{Hz}.
\end{equation}
In this process, each source of the model set was furnished with its own spectral distribution function $g:\mathbb{R}\rightarrow [0,1]$,
\begin{equation}
	g(f)=\exp\left(-\frac{1}{2}\frac{(f - f_c)^2}{f_w^2}\right),
\end{equation}
where the center frequency $f_c$ was ranging uniformly from $4\cdot\Delta_f$ to $35\cdot\Delta_f$, and the frequency width parameter $f_w$ was chosen between $\Delta_f/2$ and $64\cdot \Delta_f$, again, uniformly at random.

Finally, we included an arbitrary pressure field of degree $l_{max}=15$ (incoming towards the origin) to model ambient sound. The upper bound $u:\left\{0,\ldots,l_{max}\right\}\rightarrow \mathbb{R}$ for the magnitude of the corresponding randomly chosen complex coefficients is given by
\begin{equation}
	u(l)=u_0\exp\left(- (l_{max} + 1) \cdot \frac{(l + 1)^2 - 1}{(l_{max} + 1)^2 - 1}\right),
\end{equation}
where $u_0$ is at least $10\,\textrm{dB}$ below the model sound velocity level.

\section{Machine learning model}\label{m-l-model}

\subsection{Generative Adversarial Networks}\label{GANs}

The machine learning model we present below is based on a GAN architecture (see \cite{Goodfellow2014}), where, when training the model, a pass through the learning loop can be interpreted as a round of a zero-sum game in the sense of game theory. Here, the two players, generator and discriminator, confront each other and aim to optimize their respective objective functions. More precisely, the generator and discriminator are artificial neural networks whose parameters are optimized according to loss functions derived from their respective objective functions.

In general, the generator $	G:\mathcal{Z}\rightarrow\mathcal{B}$ is a mapping between spaces $\mathcal{Z}$ and $\mathcal{B}$, where $\mathcal{B}$, on the one hand, contains the set $\mathcal{X}\subset \mathcal{B}$ of training data (also referred to as real data) and, on the other hand, defines what is considered to be achievable through generation: the elements of the image $G(\mathcal{Z})$ are called the fake data generated by $G$ from $\mathcal{Z}$. For example, in what follows, $\mathcal{B}=\mathbb{C}^{48\times48\times16}$, therefore, it contains the cross-spectral matrices built from our sound pressure simulations in \ref{acousticModels}.

Now, the discriminator $D:\mathcal{B}\rightarrow [0,1]$ is a mapping from $\mathcal{B}$ to the unit interval. In each pass through the learning loop, a finite set $X\subseteq \mathcal{X}$ of training data is drawn randomly (this is also called mini-batching approach \cite{Bottou1998}) and complemented by a finite set $Z$ of realizations of a random variable with a fixed probability distribution taking values in $\mathcal{Z}$. The goal of the discriminator is now to distinguish between the real data $X$ and the fake data $G(Z)$ generated from $Z$. More specifically, $D$ is optimized according to the loss function
\begin{equation}\label{loss_disc}
	\mathcal{L}_D=-\frac{1}{\# X}\sum_{x\in X}\log( D(x) )-\frac{1}{\# Z}\sum_{z\in Z}\log( 1 - D(G(z)) ),
\end{equation}
where $\log$ denotes the natural logarithm. On the opposite side, the objective of the generator is to make the fake data $G(Z)$ generated from $Z$ appear as real as possible from the discriminator's point of view. This is achieved by optimizing the parameters of $G$ using the loss function
\begin{equation}\label{loss_gen}
	\mathcal{L}_G=-\frac{1}{\# Z}\sum_{z\in Z}\log( D(G(z)) ).
\end{equation}

In summary, through this adversarial process, both networks are optimized to improve their respective capabilities in processing the training data (see \cite{Salimans2016}).

\subsection{Complex model building blocks}\label{complexParts}

As the ability to take into account phase information will be crucial when working with cross-spectral matrices, all building blocks of the deep neural networks that follow will be real representations of complexifications of their traditional real-valued counterparts (see \cite{Trabelsi2018}): Suppose $L(p)\in\mathbb{R}^{N\times M}$ is a real matrix representation of a linear network operation, where $p\in\mathbb{R}^K$ is the corresponding vector of learnable parameters. Then, simply by switching to the field of complex numbers, we deduce that $L(p_r)+i\,L(p_i)\in\mathbb{C}^{N\times M}$ for $p_r,p_i\in\mathbb{R}^K$ satisfies 
\begin{align}
(L(p_r)+i\,L(p_i))(x+i\,y)=&(L(p_r)x-L(p_i)y)\\
&+i\,(L(p_r)y+L(p_i)x)
\end{align}
for all $x,y\in\mathbb{R}^M$. Therefore, we will call
\begin{equation}
	\left(\begin{matrix}
		L(p_r) & -L(p_i)\\
		L(p_i) & L(p_r)
	\end{matrix}\right)\in\mathbb{R}^{2N\times 2M}
\end{equation}
a real matrix representation of the complexification of the network operation of $L(p)$.

In addition to linear network operations, we make use of two complex phase-preserving activation functions in order to build networks that represent non-linear complex functions. Firstly, we employ a so-called modified ReLU activation function $F_{mReLU}:\mathbb{C}\rightarrow \mathbb{C}$ with bias $b\in\mathbb{R}$, which is defined as follows (see \cite{Arjovsky2016}):
\begin{align}
	F_{mReLU}(z)&=ReLU(|z|+b)\frac{z}{|z|}\\
	&=\begin{cases}(|z| + b)\frac{z}{|z|}& \text{if}\ |z| + b\geq 0,\\
		0& \text{otherwise}.
		\end{cases}
\end{align}
And secondly, we apply a leaky variant $F_{lCard}:\mathbb{C}\rightarrow \mathbb{C}$,
\begin{align}
	F_{lCard}(z)&=\frac{1}{2}\left((1 + \alpha)+\cos(arg(z))\right)z,\quad \alpha > 0,
\end{align}
of the complex cardioid activation function of \cite{Virtue2017}. The absolute value of the latter depends on the phase of $z$, whereas the former does not. 
\subsection{Models for generator and discriminator}
The generator $G=G_E\circ G_D$ we are going to employ is a composition of two convolutional neural networks, encoder $G_E$ and decoder $G_D$. The input of the encoder $G_E$ takes values of $\mathbb{C}^{48\times48\times16}$, hence, it can be applied to cross-spectral matrices built out of the simulations of \ref{acousticModels}. The kernel of the utilized (transposed) two-dimensional complex bias-free convolutions is $48\times 48$ with trivial stride. Therefore, a padding strategy was not necessary. The encoder $G_E$ is composed of one such convolutional layer $c$ and one to four complex bias-free dense layers $d$ (cf. \ref{complexParts}),
\begin{equation}\label{encoder}
	\left(\begin{matrix}48\\48\\16\end{matrix}\right)\underset{c,a}{\rightarrow} \left(\begin{matrix}1\\1\\n_{gen}\end{matrix}\right)\underset{d,a}{\rightarrow}\left(\begin{matrix}1\\1\\n_{den}\end{matrix}\right)\underset{d,a}{\rightarrow}\ldots\underset{d,a}{\rightarrow}\left(\begin{matrix}1\\1\\n_{den}\end{matrix}\right).
\end{equation}
Here, every convolution or dense layer entails an activation $a$ using one of the functions of \ref{complexParts},
\begin{equation}
	\left(\begin{matrix}d_1\\\vdots\\d_k\end{matrix}\right)
\end{equation}
is an alternative notation for $\mathbb{C}^{d_1\times\ldots\times d_k}$, the number of convolution filters is $n_{gen}\in\{32,64\}$, and the number of dense units is $n_{den}\in\{512,1024\}$. The decoder $G_D$ realizes
\begin{equation}\label{decoder}
	\left(\begin{matrix}1\\1\\n_{den}\end{matrix}\right)\underset{d,a}{\rightarrow}\ldots\underset{d,a}{\rightarrow}\left(\begin{matrix}1\\1\\n_{den}\end{matrix}\right)\underset{d,a}{\rightarrow}\left(\begin{matrix}1\\1\\n_{gen}\end{matrix}\right)\underset{c^t,a}{\rightarrow}\left(\begin{matrix}48\\48\\16\end{matrix}\right)
\end{equation}
($c^t$ denotes transposed convolution) followed by a Hermitianizing operation $H:\mathbb{C}^{48\times48\times16}\rightarrow \mathbb{C}^{48\times48\times16}$,
\begin{equation}
	(H(C))_{ijk}=\frac{1}{2}\left(C_{ijk}+C_{jik}^\ast\right).
\end{equation}

The discriminator $D$ is similar in structure to the encoder, but with removed dense layers, and the number of convolution filters now is $n_{dis}\in\{16,32\}$:
\begin{equation}\label{disc}
	\left(\begin{matrix}48\\48\\16\end{matrix}\right)\underset{c,a}{\rightarrow} \left(\begin{matrix}1\\1\\n_{dis}\end{matrix}\right)\underset{sig}{\rightarrow}[0,1].
\end{equation}
Moreover, a real sigmoid activation function $sig$ is applied the real and the imaginary parts of the convolution output (a so-called split-type A activation, see \cite{Bassey2021}), and the corresponding result is averaged over all dimensions of its real representation.
\subsection{Training set and loop}\label{loop}
Following the notation of \ref{GANs}, we now fix
\begin{equation}
	\mathcal{Z}=\mathbb{C}^{48\times48\times16},\quad \mathcal{B}=\mathbb{C}^{48\times48\times16}. 
\end{equation}
Moreover, we decide on a transformation rule the generator $G$ is intended to learn (cf.\ \ref{trafoTasks}). More precisely, we decide on sets $\mathcal{X}\subset\mathcal{B}$ and $\mathcal{Y}\subset\mathcal{B}$ and a map $f:\mathcal{X}\rightarrow\mathcal{Y}$. For example, in the most trivial case, $G$ could be trained to work as an auto-encoder on a fixed set $\mathcal{X}\subset\mathcal{B}$ by setting $\mathcal{Y}=\mathcal{X}$ and $f=\mathrm{Id}$.

In each pass through the training loop, $z_{x_i}\in Z_X$ and $z_{y_i}\in Z_{Y}$ is generated out of $x_i\in X$ and $y_i=f(x_i)\in Y=f(X)$ (with $X$ being a mini-batch drawn randomly from $\mathcal{X}$, and $Z=Z_X$, see \ref{GANs}) by adding noise, respectively. Based on \ref{GANs}, we add another term to $\mathcal{L}_G$ to integrate the transformation rule intended for $G$,
\begin{align}\label{loss_gen_gan}
	\mathcal{L}_G=&-\frac{1}{N}\sum_{i=1}^{N}\log(D(G(z_{x_i})))\\\label{loss_gen_trafo}
	&+\frac{\lambda}{2N}\sum_{i=1}^{N} \varepsilon(y_i,G(z_{x_i}))+\varepsilon(y_i,G(z_{y_i})).
\end{align}
Here, $N=\#X$, $\lambda>0$,
\begin{equation}\label{distance}
	\varepsilon(a,b)=\frac{1}{K}\sum_{k=1}^{K}d(\pi_k(a),\pi_k(b)),
\end{equation}
\begin{align}
	d(m_a,m_b)=&\kappa\left(1-\frac{tr(m_a\cdot m_b)}{\|m_a\|\|m_b\|}\right)\\
	&+(1-\kappa)\left|\|m_a\|-\|m_b\|\right|,
\end{align}
$K=16$, $\kappa = 9/10$. Moreover, $\|\cdot\|$ denotes the Frobenius norm and $\pi_k:\mathbb{C}^{48\times48\times16}\rightarrow \mathbb{C}^{48\times48}$ is the projection onto the $k$-th component,
\begin{equation}
	\pi_k(C)=(C_{ijk})_{i,j=1,\ldots,48}.
\end{equation}
To summarize \eqref{loss_gen_gan} and \eqref{loss_gen_trafo}, $G$ is sought to be a denoising opponent to the discriminator that, on the one hand, realizes the map $f$ on elements of $\mathcal{X}$, and, on the other hand, is an auto-encoder for the elements of $\mathcal{Y}$.
\section{Results}\label{results}

\subsection{Hyperparameter optimization}\label{hpo}

Some of the parameters involved in the training process were chosen based on preliminary experiments or computational limitations. For example, we fixed the mini-batch size to be $N=16$, and the size of the training data set was chosen as $\# \mathcal{X}=2560$. Moreover, each component $\pi_k(C)$ of $C\in\mathcal{X},\mathcal{Y}$ was normalized with respect to its Frobenius norm as a preprocessing step after the pressure simulations of \ref{acousticModels}. As a consequence, we were able to balance \eqref{loss_gen_gan} and \eqref{loss_gen_trafo} in terms of their magnitude by setting $\lambda=200$. For the stochastic gradient descent of generator and discriminator, we used Adam optimizers with $\beta_1=0.5$, $\beta_2=0.999$, $\epsilon=10^{-7}$ without exponential moving average (see \cite{Kingma2015}). 

The remaining parameters were chosen in a hyperparameter optimization. As metric to assess the quality of this optimization, we opted for the average accuracy
\begin{equation}\label{acc}
	g_{acc}(G)=1-\frac{1}{\# \mathcal{X}_{test}}\sum_{x\in\mathcal{X}_{test}} \varepsilon(f(x),G(x))
\end{equation}
on a test data set $\mathcal{X}_{test}$ of $512$ elements disjoint to $\mathcal{X}$ using our weighted distance function $\varepsilon$ (see \eqref{distance}) that utilizes the correlation matrix distance of \cite{Herdin2005}. The $512$ parameter combinations that were tested included the number of convolution filters in the generator $n_{gen}\in\{32,64\}$ and the discriminator $n_{dis}\in\{16,32\}$, the number of dense units $n_{den}\in\{512,1024\}$ and dense layers $n_{lay}\in\{1,2,3,4\}$ in the encoder and decoder, the learning rates $l_r\in\{2\cdot10^{-4},2\cdot10^{-5}\}$ of the Adam optimizers for generator and discriminator, and the used activation function: $F_{mReLU}$ with $b\in\{-1/8,-1/4\}$ or $F_{lCard}$ with $\alpha\in\{0,1/2\}$. The combination
\begin{align}
	n_{gen}&=64,\quad n_{dis}=16,\quad n_{den}=512,\\
	n_{lay}&=1,\quad l_r^{gen}=l_r^{dis}=2\cdot10^{-5}
\end{align}
together with $F_{lCard}$, $\alpha=1/2$ yielded the maximum value of $g_{acc}(G)=0.9866$ after $100$ epochs of training using $\mathcal{Y}=\mathcal{X}$ and $f=\mathrm{Id}$ (more specifically, task 1) in \ref{trafoTasks}).

\subsection{Transformation tasks}\label{trafoTasks}

After hyperparameter optimization, we investigated $5$ transformation tasks $G$ was intended to learn. For each of these tasks, we started by selecting a set $\mathcal{M}=\{m_1,\ldots,m_M\}$ of $M=2560$ models from our model set we presented in \ref{acousticModels}. We then simulated each of these models $m_i$ in two different complexities creating $x_i\in \mathbb{C}^{48\times48\times16}$ and $y_i\in \mathbb{C}^{48\times48\times16}$. The latter built up $\mathcal{X}$ and $\mathcal{Y}$, respectively, and we set $f(x_i)=y_i$. We considered the following pairings:
\begin{itemize}
	\item[1)] Auto-encoder pairing, $x_i=y_i$
	\item[2)] $x_i$ exhibits ambient sound, $y_i$ does not
	\item[3)] $x_i$ exhibits reflections, $y_i$ does not
	\item[4)] $x_i$ exhibits directivity, $y_i$ does not
	\item[5)] $x_i$ exhibits directivity, reflections and ambient sound, $y_i$ none of these
\end{itemize}
If not stated otherwise here, a model was simulated with monopole sources, without reflections and with no ambient sound present.

\begin{figure}[htbp]
\centerline{\includegraphics[width=0.7\columnwidth]{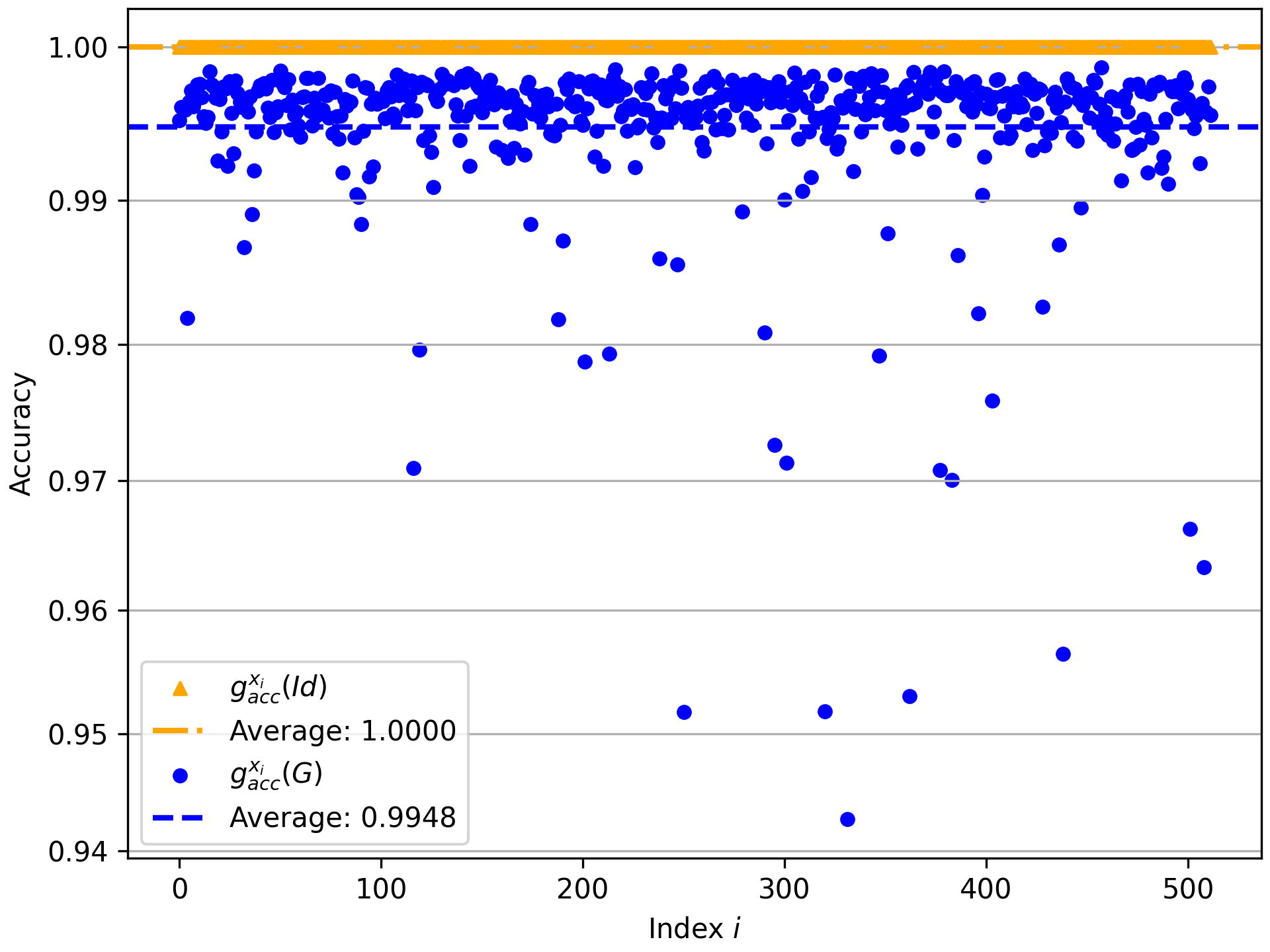}}
\caption{Accuracy scatter plot for transformation task 1) (auto-encoder)}
\label{task1}
\end{figure}

\begin{figure}[htbp]
	\centerline{\includegraphics[width=0.7\columnwidth]{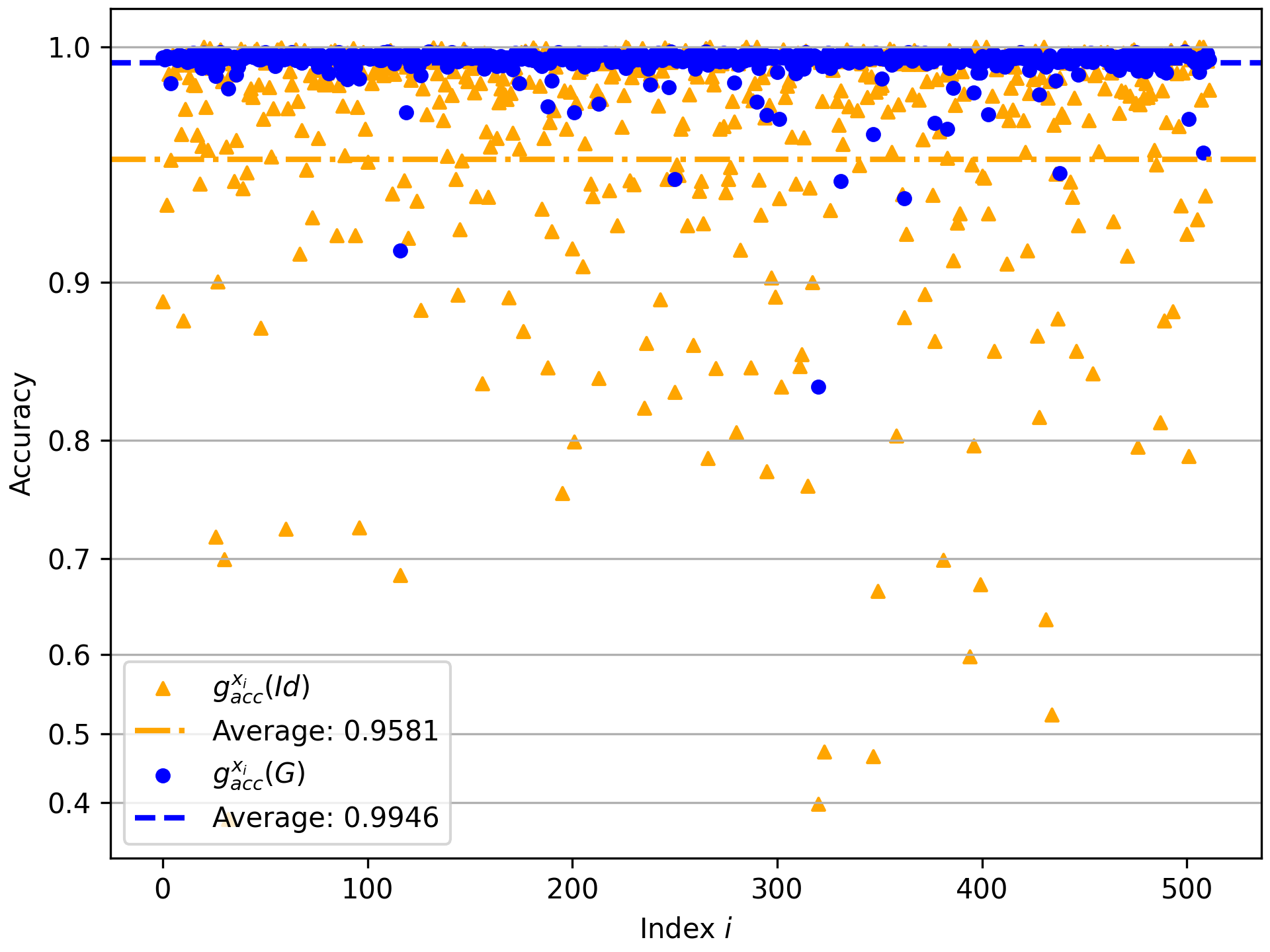}}
	\caption{Accuracy scatter plot for transformation task 2) (ambient sound)}
	\label{task2}
\end{figure}

\begin{figure}[htbp]
	\centerline{\includegraphics[width=0.7\columnwidth]{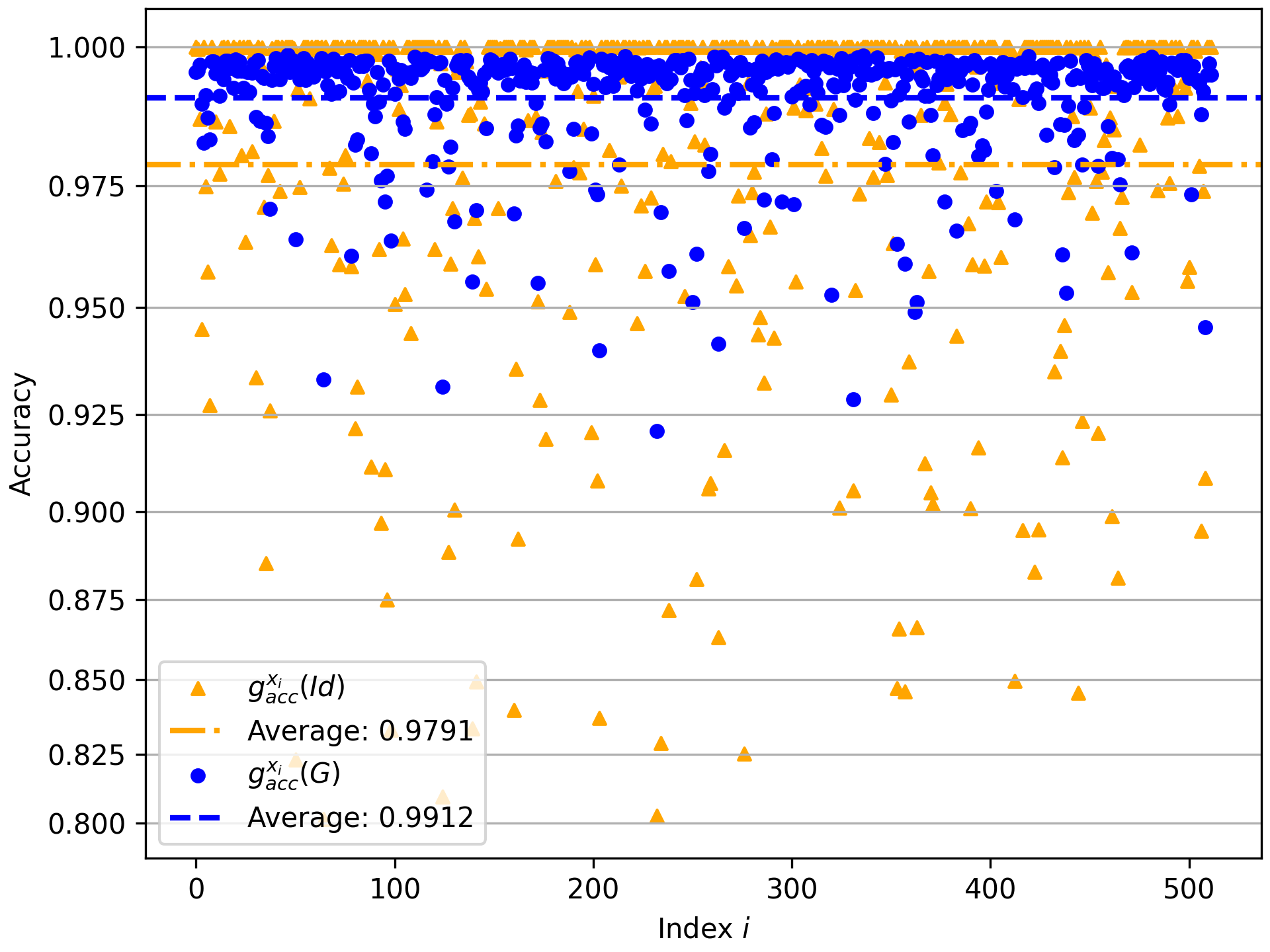}}
	\caption{Accuracy scatter plot for transformation task 3) (reflections)}
	\label{task3}
\end{figure}

\begin{figure}[htbp]
	\centerline{\includegraphics[width=0.7\columnwidth]{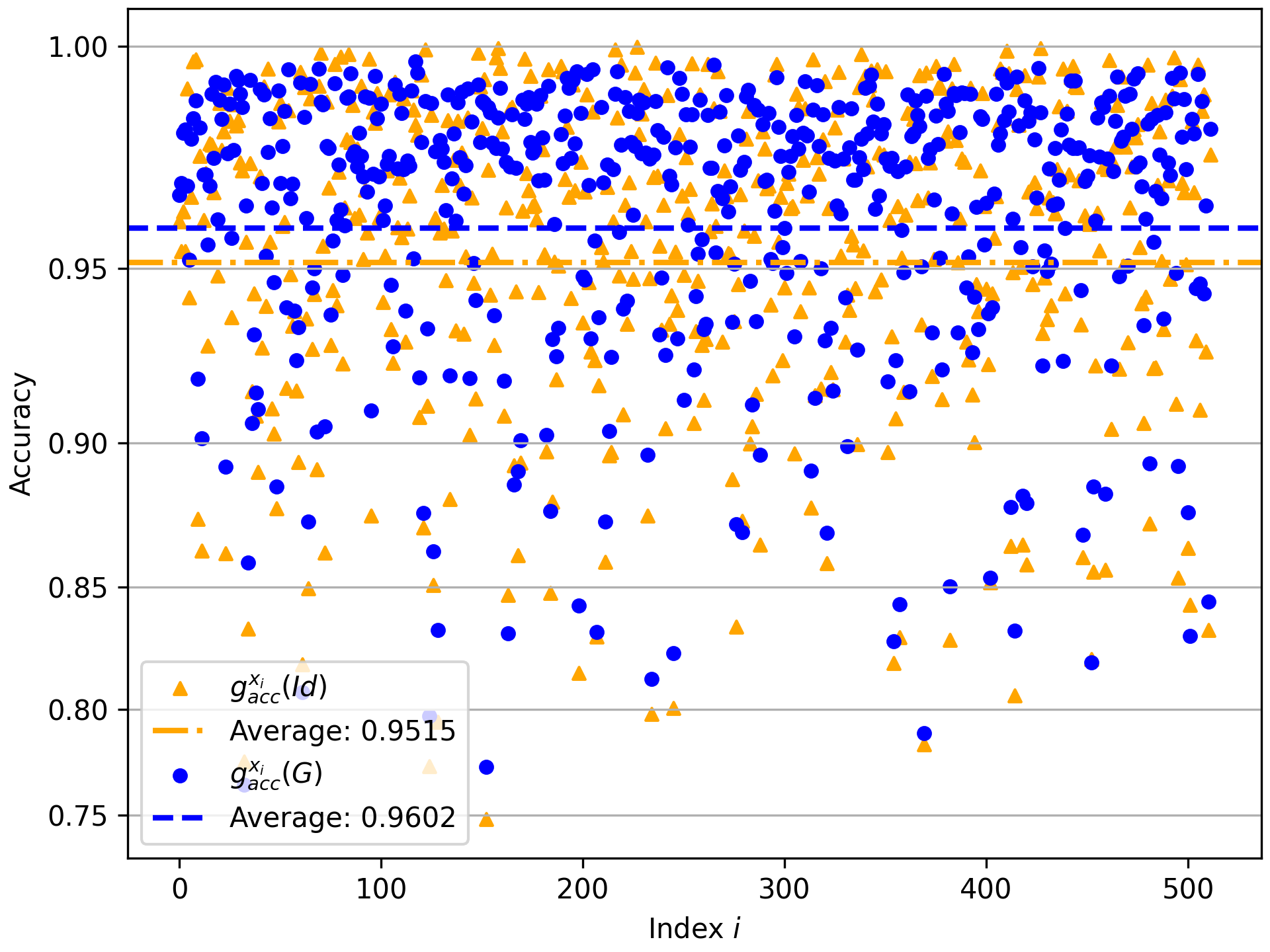}}
	\caption{Accuracy scatter plot for transformation task 4) (directivity)}
	\label{task4}
\end{figure}

\begin{figure}[htbp]
	\centerline{\includegraphics[width=0.7\columnwidth]{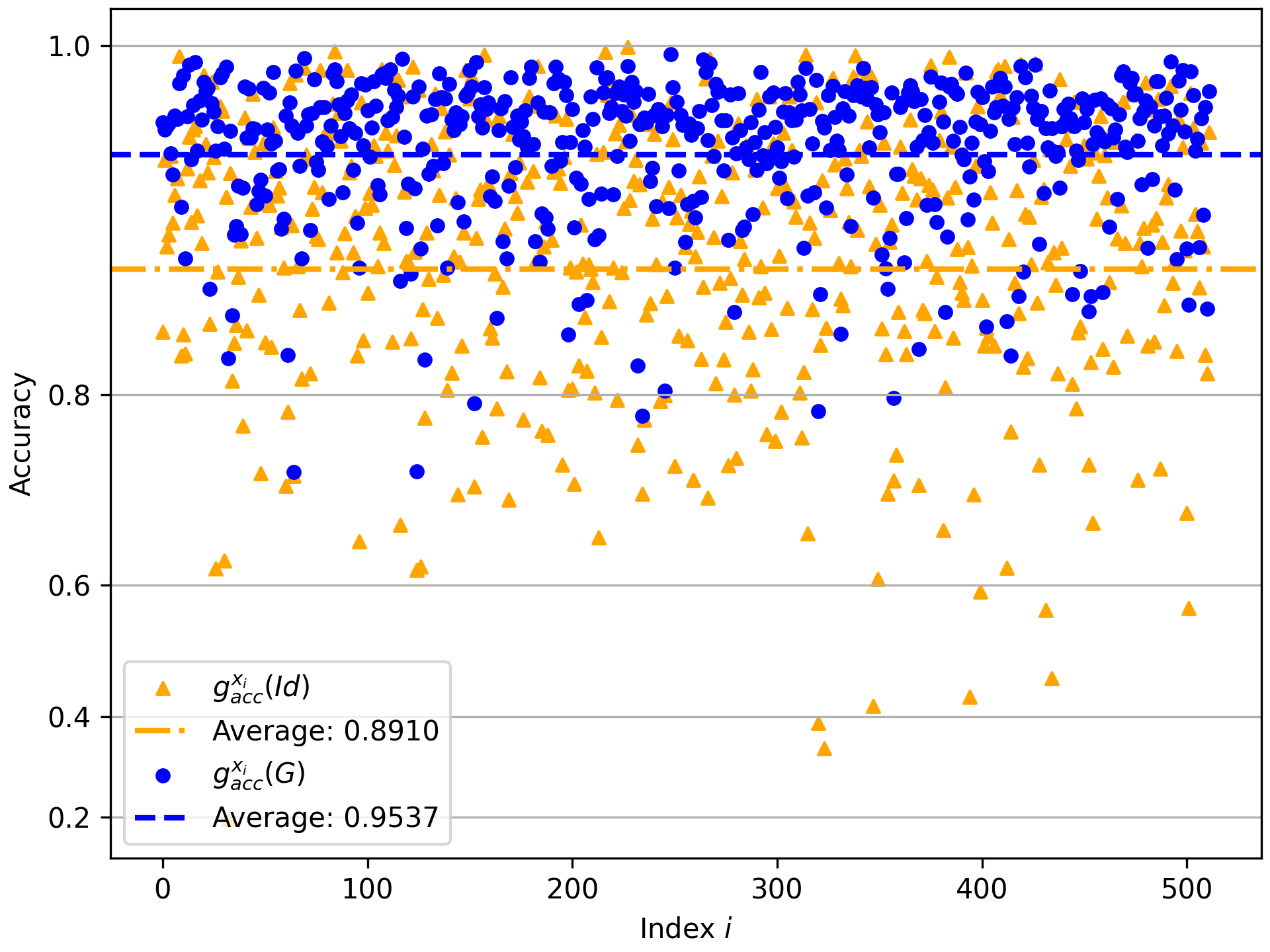}}
	\caption{Accuracy scatter plot for transformation task 5) (directivity, reflections and ambient sound)}
	\label{task5}
\end{figure}

After $1000$ epochs of training, we tested the resulting generator $G$ by reiterating the previous steps for a model set $\mathcal{M}_{test}=\{m_1,\ldots,m_{M_{test}}\}$ disjoint to $\mathcal{M}$, $M_{test}=512$. For each element $x$ of the corresponding set $\mathcal{X}_{test}$, we evaluated
\begin{equation}
	g_{acc}^x(G)=1-\varepsilon(f(x),G(x)),
\end{equation}
and $g_{acc}(G)$ (see \eqref{acc}) is simply the average of these results,
\begin{equation}
	g_{acc}(G)=\frac{1}{\# \mathcal{X}_{test}}\sum_{x\in\mathcal{X}_{test}}g_{acc}^x(G).
\end{equation}
For example, the results of the auto-encoder transformation task 1) can be found in fig.\ \ref{task1}. The average accuracy was $g_{acc}(G)=0.9948$ in this case serving as baseline, as the corresponding results of all other cases are expected to be lower or equal.
In addition, we place these values side-by-side with $g_{acc}^x(\text{Id})=1-\varepsilon(f(x),x)$ and its average $g_{acc}(\text{Id})$, respectively, in order to quantify the initial situation in comparison for all transformation tasks (see figs. \ref{task1}-\ref{task5}).

The lowest resulting $g_{acc}(G)$ values were achieved in cases 4) and 5) where the corresponding transformation task included to remove the directivity information from the data. This could be expected as the microphone array used in the simulations was only of small aperture compared to size of the total acoustic scene. Moreover, when analyzing $g_{acc}(G)$ as a function of learning epoch, it became clear, that the training progress was much slower and did not converge within $1000$ epochs of training in the cases 4) and 5). Studying fig. \ref{task3} more carefully, we realize that there are many models $m_i\in\mathcal{M}_{test}$ with $\varepsilon(f(x_i),x_i)=0$ for transformation task 3). This is due to the fact that although a reflection plane was present, its position did not render a reflection possible.

\end{document}